\documentclass[12pt,preprint]{aastex}

\begin{document}

\title{The Stellar Populations of the Cetus Dwarf Spheroidal 
Galaxy\altaffilmark{1}}

\def\lea{\mathrel{<\kern-1.0em\lower0.9ex\hbox{$\sim$}}}
\def\gea{\mathrel{>\kern-1.0em\lower0.9ex\hbox{$\sim$}}}

\author{A. Sarajedini\altaffilmark{2}, E. K. Grebel\altaffilmark{3}, 
A. E. Dolphin\altaffilmark{4}, P. Seitzer\altaffilmark{5}, 
D. Geisler\altaffilmark{6}, P. Guhathakurta\altaffilmark{7,8}, 
P. W. Hodge\altaffilmark{9}, I. D. Karachentsev\altaffilmark{10}, 
V. E. Karachentseva\altaffilmark{11}, and 
M. E. Sharina\altaffilmark{10}}

\altaffiltext{1}{Based on observations with the NASA/ESA Hubble Space 
Telescope obtained at the Space Telescope Science Institute, which is 
operated by the Association of Universities for Research in Astronomy, 
Incorporated, under NASA contract NAS5-26555.}

\altaffiltext{2}{Department of Astronomy, University of Florida, P. O. Box 
112055, Gainesville, FL 32611, USA; email: ata@astro.ufl.edu}

\altaffiltext{3}{Max-Planck-Institut f\"{u}r Astronomie, K\"{o}nigstuhl 17, 
D-69117 Heidelberg, Germany; email: grebel@mpia-hd.mpg.de}

\altaffiltext{4}{Kitt Peak National Observatory, National Optical Astronomy 
Observatory, P. O. Box 26732, Tucson, AZ 85726, USA; email: 
adolphin@noao.edu}

\altaffiltext{5}{Department of Astronomy, University of Michigan, 830 Dennison 
Building, Ann Arbor, MI 48109, USA; email: seitzer@louth.astro.lsa.umich.edu}

\altaffiltext{6}{Departamento de Fisica, Grupo de Astronomia, Universidad de 
Concepcion, Casilla 160-C, Concepcion, Chile; email: doug@kukita.cfm.udec.cl}

\altaffiltext{7}{UCO/Lick Observatory, University of California, Santa Cruz, CA 
95064, USA; email: raja@ucolick.org}

\altaffiltext{8}{Alfred P. Sloan Research Fellow}

\altaffiltext{9}{Department of Astronomy, University of Washington, 
P. O. Box 351580, Seattle, WA 98195, USA; email: 
hodge@astro.washington.edu}

\altaffiltext{10}{Special Astrophysical Observatory, Russian Academy 
of Sciences, N. Arkhys, KChR, 369167, Russia; email: ikar@luna.sao.ru}

\altaffiltext{11}{Astronomical Observatory of Kiev University, 04053, 
Observatorna 3, Kiev, Ukraine; email: vkarach@aoku.freenet.kiev.ua}

\begin{abstract}
We present Hubble Space Telescope Wide Field Planetary Camera 2 
photometry in the V and I passbands of the recently discovered Local 
Group dwarf spheroidal galaxy in Cetus. Our color-magnitude diagram 
extends from above the first ascent red giant branch (RGB) tip to 
approximately half a magnitude below the horizontal branch (HB). 
Adopting a reddening of $E(B-V) = 0.03$, the magnitude of the RGB
tip yields a distance modulus of 
$(m-M)_{0} = 24.46 \pm 0.14$. After applying the reddening and
distance modulus, we have utilized the color 
distribution of RGB stars to determine a mean metal abundance of 
$[Fe/H] = -1.7$ on the Zinn \& West scale with an intrinsic 
internal abundance dispersion of $\sim$0.2 dex. An indirect calculation 
of the HB morphology of Cetus based on the mean dereddened HB color
yields $(B-R)/(B+V+R) = -0.91 \pm 0.09$, which represents an HB that 
is redder than what can be attributed solely to Cetus' metal abundance. 
As such, Cetus suffers from the `second parameter
effect' in which another parameter besides metallicity is controlling 
the HB morphology. If we adopt the conventional `age hypothesis' 
explanation for the second parameter effect, then this implies that 
Cetus is 2-3 Gyr younger than Galactic globular clusters at its 
metallicity.
\end{abstract}


\section{Introduction}

It is truly remarkable that dwarf galaxy members of the Local Group are 
still being uncovered over 60 years after the discovery of the first 
such system, Sculptor, by Shapley (1938). Ongoing efforts to search 
for these systems have led to our current census which includes 32
dwarf galaxies in the Local Group (Grebel 2000) distributed over a 
roughly 4 Mpc$^{3}$ volume. These searches have primarily been 
fueled by the realization that dwarfs may represent the building 
blocks of much larger galaxies such as spirals and ellipticals 
(e.g. C\^{o}t\'{e} et al. 2000). Furthermore, there is the question of 
whether environmental effects play a role in the formation and 
evolution of the dwarf systems themselves. As such, understanding 
their properties is 
likely to shed light on the process of galaxy formation in general. 

Within this framework, the Hubble Space Telescope (HST) cycle 8 and 9 
programs entitled `A Snapshot Survey of Probable Nearby Galaxies'
(GO-8192 and GO-8601, PI:Seitzer) were initiated to improve our 
understanding of nearby dwarf galaxies - their numbers, spatial 
distribution, structural properties, and stellar populations, among 
other things (Seitzer et al. 2001). Previous papers reporting the 
results of our survey include Dolphin et al. (2001) and Karachentsev et 
al. (1999; 2000a; 2000b). 

In the present work, we turn our attention to the 
Cetus dwarf spheroidal (dSph) galaxy, which was discovered by Whiting,
Hau, \& Irwin (1999; hereafter WHI) as part of their comprehensive  
program using sky survey plates to search for low surface brightness 
dwarf galaxy candidates. It is located at 
$\alpha_{2000} = 00^{h} 26^{m} 11^{s}, 
\delta_{2000} = -11^{\circ} 02' 40''$ in a relatively isolated region 
of the Local Group.
Follow-up observations of Cetus with the Cerro Tololo 
Inter-American Observatory 1.5m telescope allowed WHI
to determine a distance modulus of $(m-M)_0 = 24.45 \pm 0.15$ 
based on the magnitude of the red giant branch (RGB) tip and a 
metallicity of $-1.9 \pm 0.2$ dex from the color of the RGB. In addition, 
from the surface brightness profile of Cetus, WHI estimated a core 
radius of $1.5 \pm 0.1$ arcmin and a tidal radius of $4.8 \pm 0.2$ 
arcmin. Subsequent work by Tolstoy et al. (2000, see also Tolstoy 
2001) presents a deep $BR$ color-magnitude diagram for Cetus based on 
observations with the ESO Very Large Telescope. Their data reveal
that Cetus possesses a predominantly red horizontal branch (HB)
with the possible presence of a small blue HB component. 

The remainder of this paper is organized as follows. The next section 
describes how the observations were obtained and the techniques 
employed to derive our color-magnitude diagram for Cetus. 
Sec. 3 presents an analysis of our results regarding the distance 
(Sec. 3.2), metallicity (3.3), and horizontal branch morphology (3.4) 
of Cetus. Finally, our conclusions are summarized in Sec. 4.

\section{Observations and Data Reduction}

The observations of Cetus were obtained on 16 Jan 2001 UT as part of 
our cycle 9 HST snapshot survey focusing on 
nearby dwarf galaxies (GO-8601, Seitzer et al. 2001). The data consist of two 
600s exposures, one each using the F606W ($\sim$V) and F814W 
($\sim$I) filters, with Cetus 
centered in the WF3 CCD. Figure 1a shows our observed WFPC2 field 
overlaid on the Digitized Sky Survey while Fig. 1b displays a 
cosmic ray cleaned mosaic of our combined F606W and F814W WFPC2 images. 

The photometric reduction of these observations has already 
been fully described by Dolphin et al. (2001). To summarize, the 
calibrated images obtained from STScI were input into the 
HSTphot package (Dolphin 2000a).
After removing cosmic rays with the HSTphot \textit{cleansep} routine, 
simultaneous photometry was performed on the F606W and F814W frames
using \textit{multiphot}, with aperture corrections made to an
aperture of radius 0.5 arcsec.  CTE corrections and calibrations were
then applied using the Dolphin (2000b) formulae, producing $VI$
photometry for all stars detected in both images.
Because of the relatively small field of the 
PC1 CCD, very few bright stars were available for the computation of 
an aperture correction, leading to larger than acceptable uncertainties
in the PC1 photometry. As a result, we have excluded the PC1 results 
from further analysis. In the case of the WF photometry, 
we estimate an uncertainty of 0.02 mag in the photometric zeropoint
based on the results of Dolphin (2000b).
In contrast to previous papers in this series, instead of setting a 
lower limit of 5 for the signal-to-noise per stellar magnitude 
measurement, we have used a lower limit of 3.5. This has been done so as 
to provide the maximum amount of information on the morphology of the 
blue horizontal branch, which is near the limit of the photometry.

\section{Results and Discussion}

\subsection{Color-Magnitude Diagrams}

Figure 2 shows our (V,V--I) and (I,V--I) 
color-magnitude diagrams (CMDs) for the Cetus dSph based on photometry 
derived from the three WF chips for 3101 stars. The variation of the photometric 
errors with magnitude are also shown. The CMDs reveal a relatively 
steep and well-populated red giant branch (RGB) along with a 
predominantly red horizontal branch (HB).  The tip of the first ascent 
RGB is located at I$\sim$20.5 with the level of the HB being at 
V$\sim$25. The CMD contains little or no indication of significant recent 
star formation in Cetus. For example, there are very few stars in the 
region where one expects to see O and B type stars (V--I $\lea$ 0)
indicative of a $\lea$1 Gyr population. In addition, there is a 
negligible number of asymptotic giant branch stars located above 
the first ascent RGB, which 
are normally associated with an intermediate-age (1$\lea$t$\lea$8 Gyr)
population.  

\subsection{Reddening and Distance}

Both the Burstein \& Heiles (1982) and Schlegel, Finkbeiner, \& 
Davis (1998) reddening maps indicate that Cetus is in a region of 
very low extinction. More specifically, the latter suggest that 
the reddening of Cetus is $E(B-V) = 0.03$, which, following the lead 
of WHI, we will also adopt in the present study along with an assumed 
error of 0.01 mag. In 
addition, given that $A_{V} = 3.1 E(B-V)$, for the HST filters used 
herein, Schlegel et al. (1998) give $A_{I} = 1.85 E(B-V)$ so that
$E(V-I) = 1.25 E(B-V)$.

Since the location of the first ascent RGB-tip (TRGB) appears to
be well determined in our CMD, we will obtain an estimate of the 
distance using I(TRGB). Our RGB cumulative luminosity function is shown in 
Fig. 3 to which we have applied a slope finding algorithm that isolates 
the TRGB (dashed line). This technique yields $I(TRGB) = 20.50 \pm 0.10$, 
in good accord with the WHI value. Assuming 
an absolute magnitude of $M_I(TRGB) = -4.05 \pm 0.10$ 
(Da Costa \& Armandroff 1990, hereafter DCA; 
Sakai, Madore, \& Freedman 1996; Bellazzini, Ferraro, 
\& Pancino 2001), this gives 
$(m-M)_{0} = 24.46 \pm 0.14$, which is in good accord with the WHI 
value of $(m-M)_{0} = 24.45 \pm 0.15$; to minimize confusion, we will 
adopt their distance modulus in the subsequent analysis.

We note in passing that the HB can in principle be used as a distance 
indicator as well. However, since we have no information on the presence of 
RR Lyrae variables in Cetus, the use of a relation between $M_{V}(RR)$ 
and $[Fe/H]$ may produce spurious results. Carretta et al. (2000) 
provide a calibration of the mean magnitude of the HB [$M_{V}(HB)$] 
in terms of metallicity, but the only red HB clusters in their 
sample, 47 Tuc and NGC 362, have globular cluster-like ages. Yet, it 
is known that {\it both} age and metallicity affect the luminosity of 
a purely red HB (a.k.a. red clump; Sarajedini 1999; Girardi \& Salaris 
2001). Therefore, if we assume that the mean age of the Cetus population 
producing its red HB is not significantly different from those of 
47 Tuc and NGC 
362, then we can use Equation 6 of Carretta et al. (2000) and our mean 
metal abundance (Sec. 3.3) to calculate $M_{V}(HB) = 0.51 \pm 0.08$.
Fitting a Gaussian to our $V$ luminosity function, we find
$V(HB) = 25.01 \pm 0.02$ (sec. 3.4), which produces a distance modulus of
$(m-M)_{0} = 24.41 \pm 0.09$, consistent with the TRGB distance obtained
above. 

\subsection{Metallicity}

At this point, since we have estimates for the reddening and distance 
of Cetus, it is a relatively simple matter to compare our RGB 
photometry to globular cluster RGB sequences.
The two panels of Fig. 4 show our Cetus photometry in the 
($M_{V},(V-I)_0$) and ($M_{I},(V-I)_0$) planes along with the
empirical RGBs of Saviane et al. (2000) for metallicities of
$[Fe/H] = -2.2, -1.6,$ and --1.2 on the Zinn \& West (1984) abundance 
scale and assuming the Lee, Demarque, \& Zinn (1990) distance scale 
(both the same as DCA). These sequences are based on their 
V,I database of Galactic globular cluster photometry and agree quite well 
with the RGBs published by DCA for M15, M2, and NGC 1851.
From this figure, the mean metal 
abundance of Cetus would appear to be somewhere between --2.2 and --1.6. 

To obtain a more quantitative assessment, we have utilized  
equations (2) through (6) in Saviane et al. (2000)
along with the relevant coefficients in their Table 5 to calculate the 
metallicity for all Cetus RGB stars with $-4.0 \leq M_{I} \leq -2.0$.
In addition, the error in metal abundance is estimated by multiplying 
the HSTPhot error in $V-I$ by d$[Fe/H]$/d$(V-I)_{0}$. The resultant 
metallicity distribution function (MDF) for stars with 
$-4.0 \leq M_{I} \leq -3.0$ is shown in Fig. 5a,
while the MDF for stars with $-3.0 < M_{I} \leq -2.0$ is included in
Fig. 5b. 

We note that 
inspection of globular cluster VI CMDs and theoretical isochrones 
suggest that the AGB merges into the RGB at $M_{I}\sim-2.0$; this 
indicates that 
our MDF should not be significantly affected by 
(bluer) AGB stars simulating an excess of metal-poor stars. We will 
now show that the characteristics of the MDF are unaffected by the 
magnitude range as long as only stars 
brighter than $M_{I}\sim-2.0$ are used.

The dashed curves in Fig. 5
are Gaussian fits to the plotted MDFs. For 107 stars with 
$-4.0 \leq M_{I} \leq -3.0$ (upper panel in Fig. 5), we find a peak 
abundance of $[Fe/H] = -1.71$ 
with a 1-$\sigma$ dispersion of 0.22 dex. Among these same stars, the 
mean $[Fe/H]$ error per star is 0.15 dex leading to an intrinsic 
1-$\sigma$ abundance spread of 0.17 dex. In the case of the 144 RGB 
stars with $-3.0 < M_{I} \leq -2.0$ (lower panel in Fig. 5), 
the peak abundance is 
$[Fe/H] = -1.65$ with a 1-$\sigma$ spread of 0.36 dex, which when 
coupled with the abundance dispersion introduced purely by the 
photometric errors of 
0.30 dex, yields an intrinsic 1-$\sigma$ metallicity dispersion of 
0.19 dex. Thus, we conclude that Cetus exhibits a mean metallicity 
of $[Fe/H] = -1.7$. In addition, we find an intrinsic 
1-$\sigma$ abundance dispersion of $\sim$0.2 dex in Cetus; 
Table 2 in Grebel (2000) suggests that metal 
abundance dispersions of this order are 
quite common among the dwarf galaxies in the Local Group. With 
high-resolution spectroscopy, even larger dispersions are being 
uncovered (e.g. Shetrone, C\^{o}t\'{e}, \& Sargent 2000).

We note that our determination of the mean metallicity of Cetus is
0.2 dex more metal-rich than that of WHI. Since we adopt the same
distance and reddening as WHI, this abundance difference 
is likely to be the result of differences in our V--I color
zeropoints. A comparison of the RGB location in the WHI CMD with our 
CMD suggests that the former is $\sim$0.05 mag bluer than the latter.
The origin of this photometric zeropoint difference is unclear.

We acknowledge that our measured values of the mean metallicity and
metallicity dispersion can be influenced by age effects, as many (or most)
dwarf spheroidals formed their stars over extended periods of time.
Based on the Girardi et al. (2000) isochrones, we estimate that
for ages greater than 10 Gyr, this effect is minimal ($\Delta$t of 5
Gyr corresponds to $\Delta [Fe/H]$ of roughly 0.1 dex in terms of RGB
color), thus leaving our metallicity measurements largely unchanged.
However, we cannot completely rule out the presence of younger
stars without observing the main sequence turnoff; if stars of age
$\sim 7$ Gyr or younger are present, the resulting effect on our
metallicity measurement is 0.2 dex or greater, also adding a large
uncertainty to our metallicity dispersion measurement.  Although this
scenario is possible, there is no {\it a priori} reason to assume
the presence of younger stars, and thus we will continue to use our
measured value of $\langle$$[Fe/H]$$\rangle$$ = -1.7$ with a 1-$\sigma$ 
dispersion of 0.2 dex.

We have also utilized the data in Table 2 of Grebel (2000) to plot the 
mean metallicity of Local Group dwarfs as a function of their absolute 
magnitude (Fig. 6a) and central surface brightness (Fig. 6b), 
excluding the Sagittarius dwarf galaxy and M32. The 
plotted data represent dSphs (crosses), dIrrs (filled circles) and 
dEs (open circles). There is a 
well-known relation between mean abundance, absolute V magnitude, and 
central surface brightness for these systems 
in the sense that more luminous galaxies 
with higher central surface brightness tend to enrich themselves in 
heavy elements to a greater degree (Caldwell et al. 1998, and 
references therein). 
The large open square represents the location of 
Cetus based on our new mean metal abundance coupled with the luminosity and 
surface brightness estimates of WHI. It is clear that Cetus occupies 
its expected location in these diagrams.

\subsection{Horizontal Branch}

As first illustrated in Fig. 2, the HB of Cetus appears to be primarily
populated redward of the RR Lyrae instability strip. As a comparison,
Fig. 7 shows our Cetus CMD compared with the fiducial sequences of the 
globular clusters Rup 106 (Sarajedini \& Layden 1997),
M3 (Johnson \& Bolte 1998) and M54 (Sarajedini \& Layden 1995). 
While the metallicity of Cetus is 
similar to those of Rup 106, 
M3 and M54 (the RGBs match), there is no question that its HB 
morphology is significantly redder than the HBs of these clusters.
We are probably missing some blue HB 
stars in Cetus (especially those that may be present along the blue 
HB tail, i.e. fainter than $M_{V}\sim +1$) due to photometric 
incompleteness, but there are unlikely to be as many blue HB stars as those 
on the red HB. 
We also point out that the (I,V--I) CMD (bottom panel
of Fig. 2) shows that the photometric incompleteness along the RGB
is not severe even at the faintest levels, i.e. the luminosity
function of RGB stars is steadily increasing at these magnitudes.
This suggests that photometric incompleteness on the HB at similar magnitude
levels is not likely to be problematic.

It has been traditional to use the 
numbers of stars blueward of the instability strip (B) and those
redward of it (R) along with stars within the strip itself 
(RR Lyrae variables, V) in order 
to construct the $(B-R)/(B+V+R)$ index (Lee et al. 1994). However, 
this approach is not practical in our case because of the lack of 
information 
on the numbers of RR Lyraes in Cetus. Instead, we adopt a different 
procedure, one that takes advantage of the peak HB color as defined 
by Buonanno et al. (1997). For clusters with metallicities between
$-1.80\leq[Fe/H]\leq-1.60$ as given in Table 1 of Buonanno et al.
(1997), Fig. 8 displays the relation between the 
peak $(B-V)_{0}$ color of Galactic globular clusters
and their $(B-R)/(B+V+R)$ values (Lee et al. 1994). The plotted 
quantities
are listed in Table 1. The peak $(B-V)_{0}$ values come from 
Buonanno et al. (1997) except for Rup 106, which is calculated 
using the photometry of Kaluzny, Krzemenski, \& Mazur (1995).

The next step is to measure the peak $(V-I)_{0}$ of the 
Cetus HB and convert it to $(B-V)_{0}$. First, we construct luminosity 
functions of our photometry in V and I to which we fit Gaussian 
distributions in order to isolate the magnitude of the HB. We find 
$V(HB) = 25.01 \pm 0.02$ and $I(HB) = 24.22 \pm 0.02$ where the 
errors represent a combination of the standard errors of the mean
and an estimate of the uncertainty in the fitting process. Subtracting these 
and including an estimated error of 0.02 mag in the photometric 
zeropoint along with our adopted Cetus reddening gives 
$(V-I)_{0} = 0.75 \pm 0.03$. We prefer this method rather than 
directly measuring the peak color of the HB for two reasons. First, 
the color distribution of the HB 
stars is more susceptible to contamination from the color peak
of the RGB stars which are slightly redward. Second, the LF 
peaks of the HB in V and I are more closely Gaussian so that fitting 
them with such a distribution is more straightforward.
A conversion from $(V-I)_0$ to $(B-V)_0$, however, is somewhat more
uncertain.  Zinn \& Barnes (1996) provide a horizontal branch color
conversion equation; however it is based on M15 and M68, whose horizontal
branches are both predominantly blue.  Thus the value we calculate 
for red HB stars
(which is what we have in Cetus) is interpolated between their blue HB
stars and RGB stars at the level of the HB.  Using their relation, we find
$(B-V)_{0} = 0.55 \pm 0.03$. Another empirical relation is given by
equations 8 and 10 of von Braun et al. (1998), based only on RGB stars;
this gives $(B-V)_{0} = 0.48 \pm 0.05$. Finally, one can use the
Kurucz (1994) model atmospheres, convolved with the Bessell (1990)
passbands and normalized with the Hamuy et al. (1992) spectrophotometric
standards, to obtain theoretical transformations; this procedure results
in a color of $(B-V)_{0} = 0.55 \pm 0.03$.  Because our use of the von Braun
et al (1998) relations is an extrapolation from RGB stars, we utilize the
other two values to measure $(B-V)_{0} = 0.55 \pm 0.03$ for the peak
color of our horizontal branch.  Using the relation plotted in Fig. 8,
this color produces a horizontal branch index $(B-R)/(B+V+R)$ of
$-0.91 \pm 0.09$. 

Based on its intermediate metallicity and red HB morphology, Cetus joins 
a long list of other Local Group dSphs that suffer from the 
second parameter effect (e.g. Harbeck et al. 2001), in much 
the same way as the Galactic globular clusters in the outer halo 
(Sarajedini, Chaboyer, \& Demarque 1997, and references therein). 
That is to say, the HB morphology of Cetus is too red for its mean metal 
abundance. Although there is still no general agreement on the cause 
of the second parameter effect, the conventional explanation is based on 
the age hypothesis in which the stellar population of Cetus would be 
younger than a typical Galactic GC. Utilizing new HB models 
published by Rey et al. (2001, see their Fig. 4), we estimate 
that this age difference amounts to some 2-3 Gyr.  

\section{Conclusions}

Using HST/WFPC2 in snapshot mode, we have constructed a deep CMD of 
the Cetus dwarf spheroidal galaxy. Based on this CMD, which extends 
from the tip of the RGB to about 0.5 mag below the HB, and an adopted 
reddening of $E(B-V) = 0.03$ based on the Schlegel et al. (1998) 
extinction maps, we are able to 
determine a number of fundamental properties for this galaxy. First, 
we exploit the tip of the first ascent RGB to derive a distance 
modulus of $(m-M)_{0} = 24.46 \pm 0.14$, which agrees very well with that of 
Whiting et al. (1999). Having corrected for the reddening and the 
distance, a comparison of the Cetus RGB with empirical sequences 
yields a mean metal abundance of $[Fe/H] = -1.7$ on the Zinn \& West 
(1984) scale with an intrinsic 
internal abundance dispersion of 0.2 dex. The HB of Cetus is 
populated predominantly redward of the RR Lyrae instability strip; we 
calculate an HB morphology index of $(B-R)/(B+V+R) = -0.91 \pm 0.09$
based on a technique that utilizes the mean 
dereddened color of the HB. As such, the mean 
metallicity and HB morphology of Cetus suggest that the stellar 
populations of this galaxy manifest the second parameter effect. If we 
accept the explanation that this is due to age, then Cetus is about 
2-3 Gyr younger in the mean than typical Galactic globulars at its 
metallicity.

\acknowledgements

Support for proposals GO-8192 and GO-8601 was provided by NASA through
a grant from the Space Telescope Science Institute, which is operated 
by the Association of Universities for Research in Astronomy, 
Incorporated, under NASA contract NAS5-26555. 
A.S. has also benefited from financial support from NSF CAREER 
grant No. AST-0094048. D. G. gratefully acknowledges financial 
support for this project received from CONICYT through FONDECYT grant 
8000002.

\begin{deluxetable}{lccc}
\tablewidth{4in}
\tablecaption{Cluster Horizontal Branch Parameters} 
\tablehead{
\colhead{Cluster} & \colhead{$[Fe/H]$} & \colhead{$(B-V)_{0}$} & 
\colhead{$(B-R)/(B+V+R)$}}
\tablecolumns{4}
\startdata
NGC 1904 & --1.69 & 0.05 & $0.89\pm0.16$ \\
NGC 4147 & --1.80 & 0.20 & $0.55\pm0.14$ \\
NGC 5272 & --1.66 & 0.30 & $0.08\pm0.04$ \\
NGC 5897 & --1.68 & --0.03 & $0.91\pm0.10$ \\
NGC 6218 & --1.61 & --0.07 & $0.92\pm0.10$ \\
NGC 6254 & --1.60 & --0.10 & $0.94\pm0.10$ \\
Pal 14   & --1.60 & 0.65 & $-1.00\pm0.08$ \\
Rup 106  & --1.80 & 0.46 & $-0.82\pm0.15$ \\
\enddata
\end{deluxetable}

\clearpage

\figcaption{(a) The outline of the Wide Field Planetary Camera 2 field on 
the Digitized Sky Survey image of Cetus showing our observed field. 
(b) A cosmic ray cleaned mosaic of our combined
F606W and F814W WFPC2 images. The compass 
marking indicates the directions of north (arrow) and east.}

\figcaption{The (V,V--I) and (I,V--I) 
color-magnitude diagrams for the Cetus dSph based on photometry 
for 3101 stars derived from the three WF chips. The right panels show 
the variation of photometric error with increasing magnitude.}

\figcaption{The cumulative luminosity function (LF) of the Cetus 
red giant branch (RGB) is shown 
by the solid line. The derivative of this LF, represented by the 
dotted line, shows a significant increase at $I= 20.5$ 
indicating the onset of the RGB tip.}

\figcaption{Distance and reddening corrected CMDs for Cetus. The solid 
lines are the Saviane et al. (2000) red giant branch sequences for 
metallicites of $[Fe/H] = -2.2$, --1.6, and --1.2 on the Zinn \& West 
(1984) abundance scale and assuming the Lee et al. (1990) distance 
scale. The mean photometric errors at each magnitude level are 
indicated.}

\figcaption{The metallicity distribution function of Cetus. The upper 
histogram uses red giant branch (RGB) stars with 
$-4.0 \leq M_{I} \leq -3.0$, while the lower histogram uses RGB stars 
with $-3.0 < M_{I} \leq -2.0$. The dashed lines represent 
Gaussian fits to these distributions as discussed in the text.}

\figcaption{The behavior of the mean metal abundance for dwarf galaxies in 
the Local Group with absolute V magnitude and central surface 
brightness in the V band. As indicated in the legend, the filled 
circles are the dwarf irregulars, the open circles are the dwarf 
ellipticals, and the crosses are the dwarf spheroidals. The location 
of Cetus is indicated by the open square.}

\figcaption{A comparison of our Cetus CMD with the fiducial sequences 
of Rup 106 (Sarajedini \& Layden 1997), 
M3 (Johnson \& Bolte 1998) and M54 (Sarajedini \& 
Layden 1995). These clusters have similar 
metallicities to Cetus and thus their red giant branches match
that of Cetus reasonably well. However, Cetus appears to have a horizontal 
branch (HB) that is significantly redder than the HBs of these globular 
clusters.}

\figcaption{The relation between the peak dereddened color of the 
horizontal branch and the $(B-R)/(B+V+R)$ HB morphology index. The 
plotted data are given in Table 1. The dashed line is the least squares 
fit to these data and has been constructed assuming the abscissa to be 
the dependent variable and the ordinate to be the independent variable.}

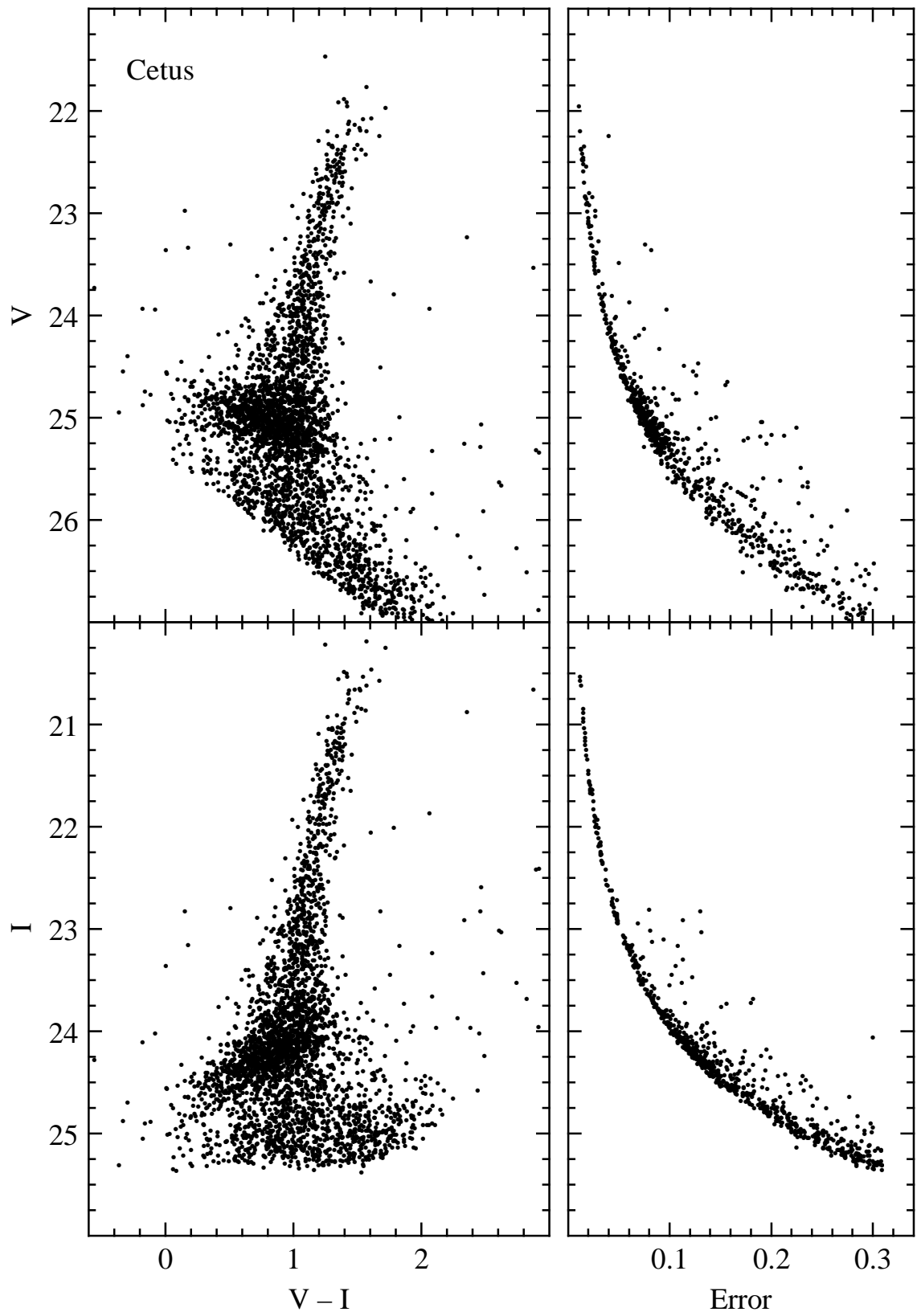
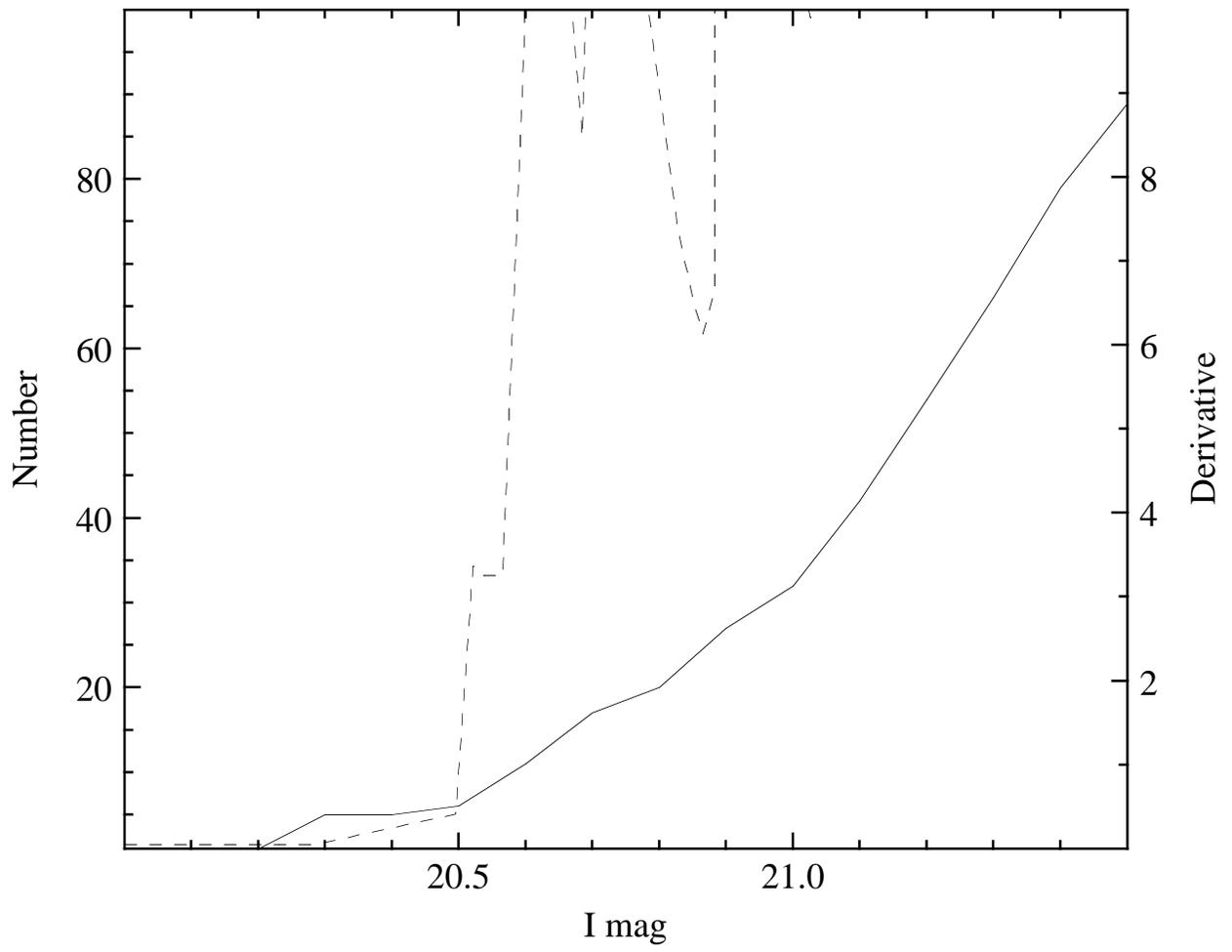
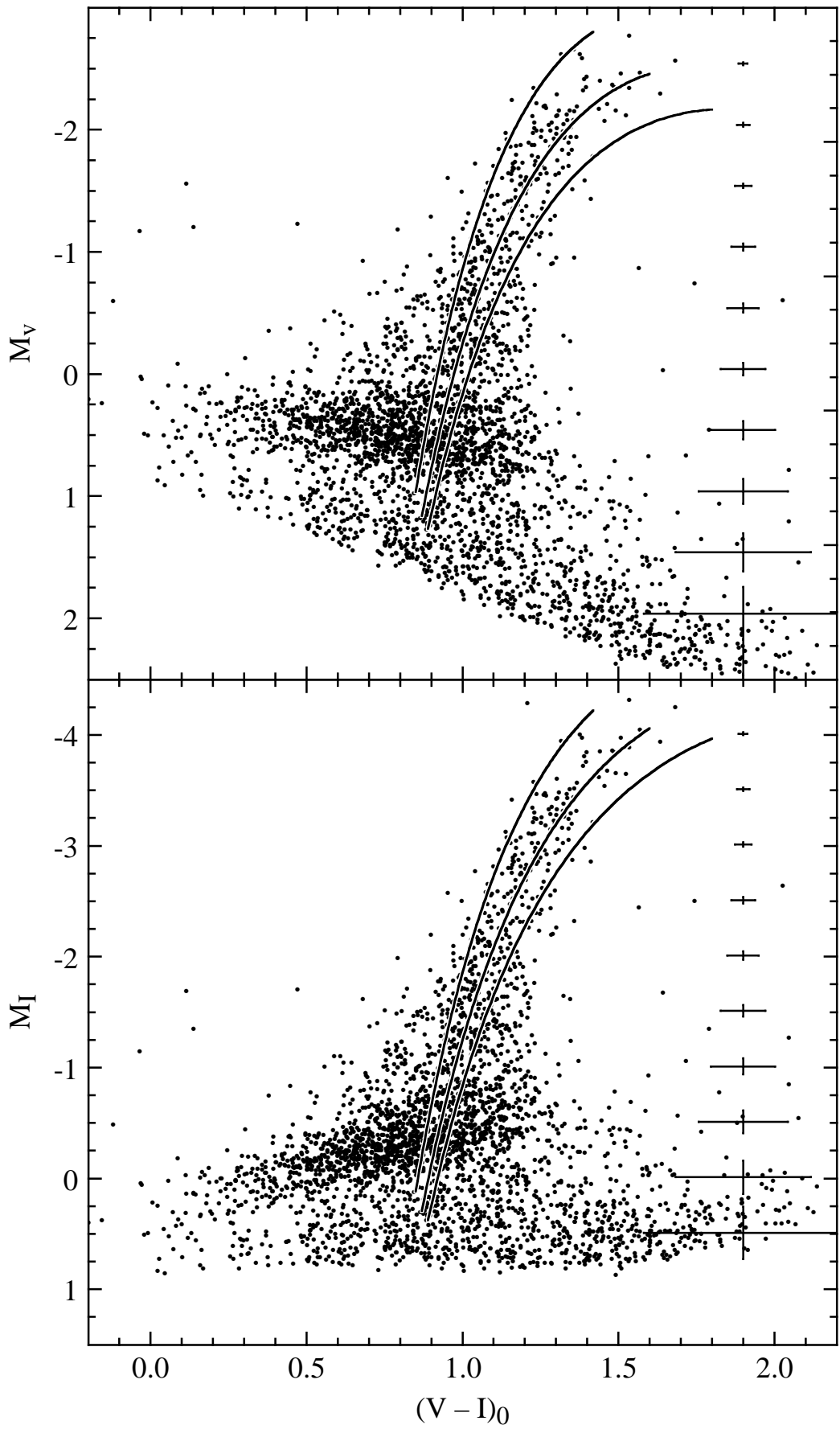
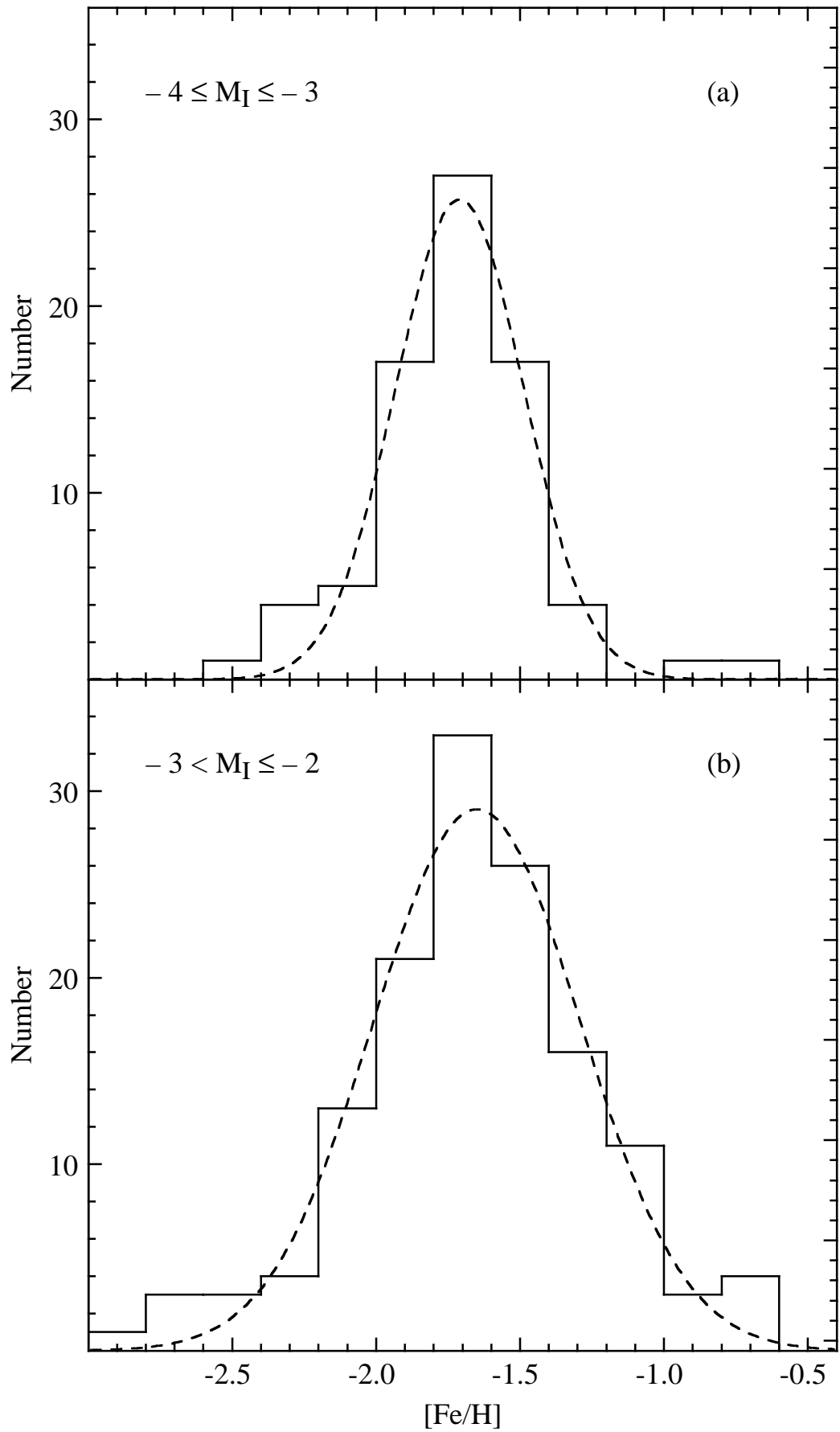
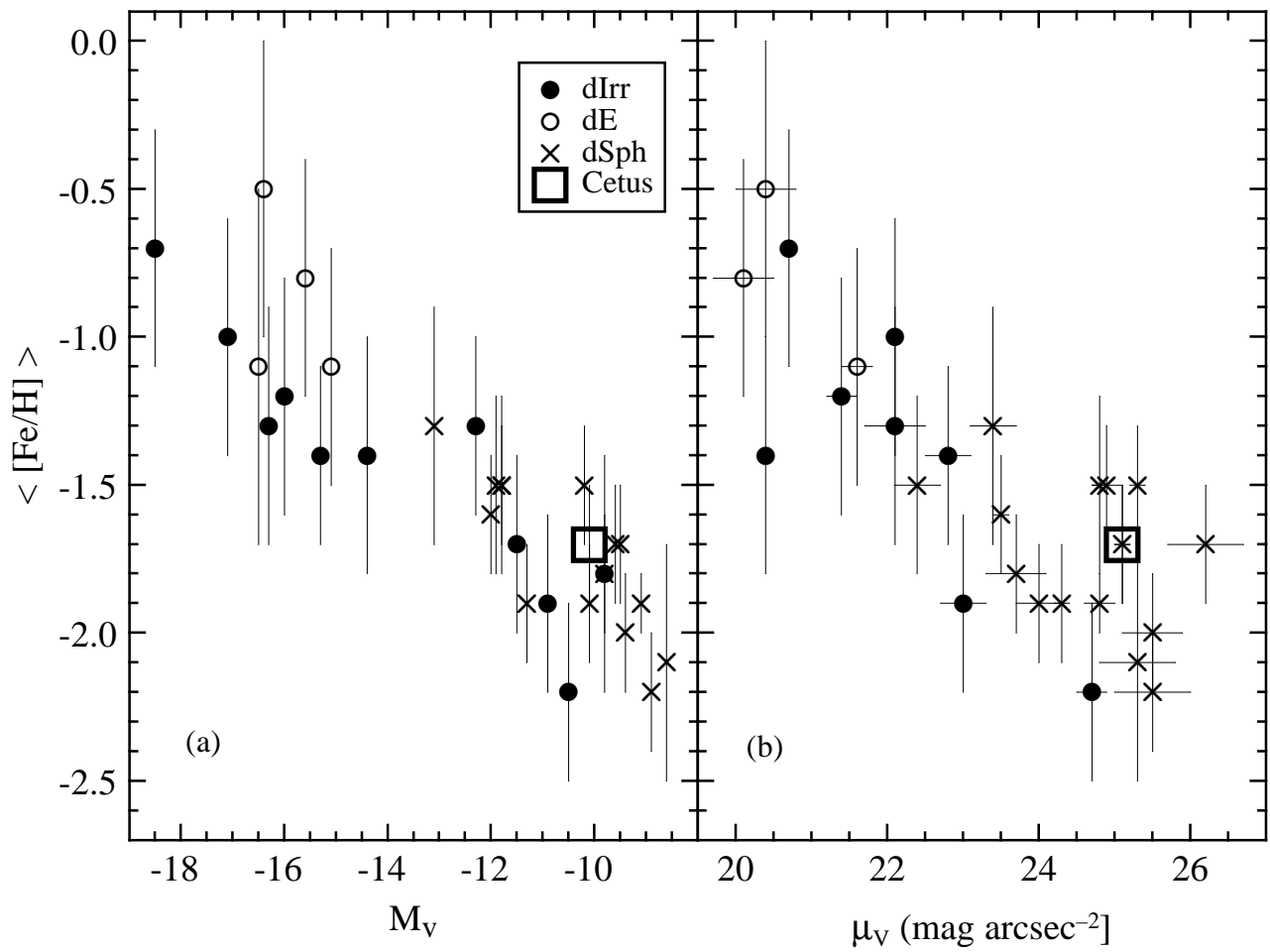
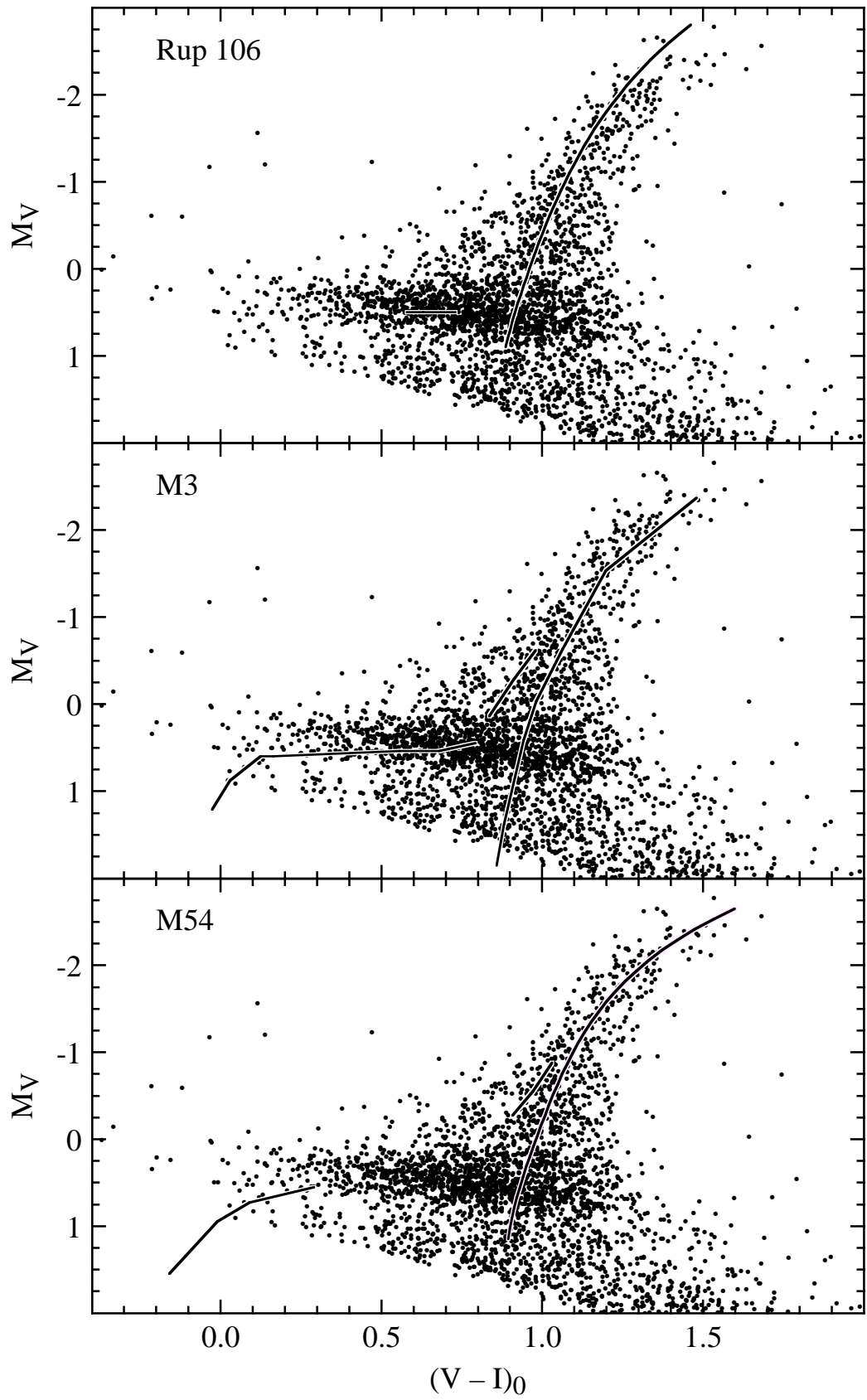
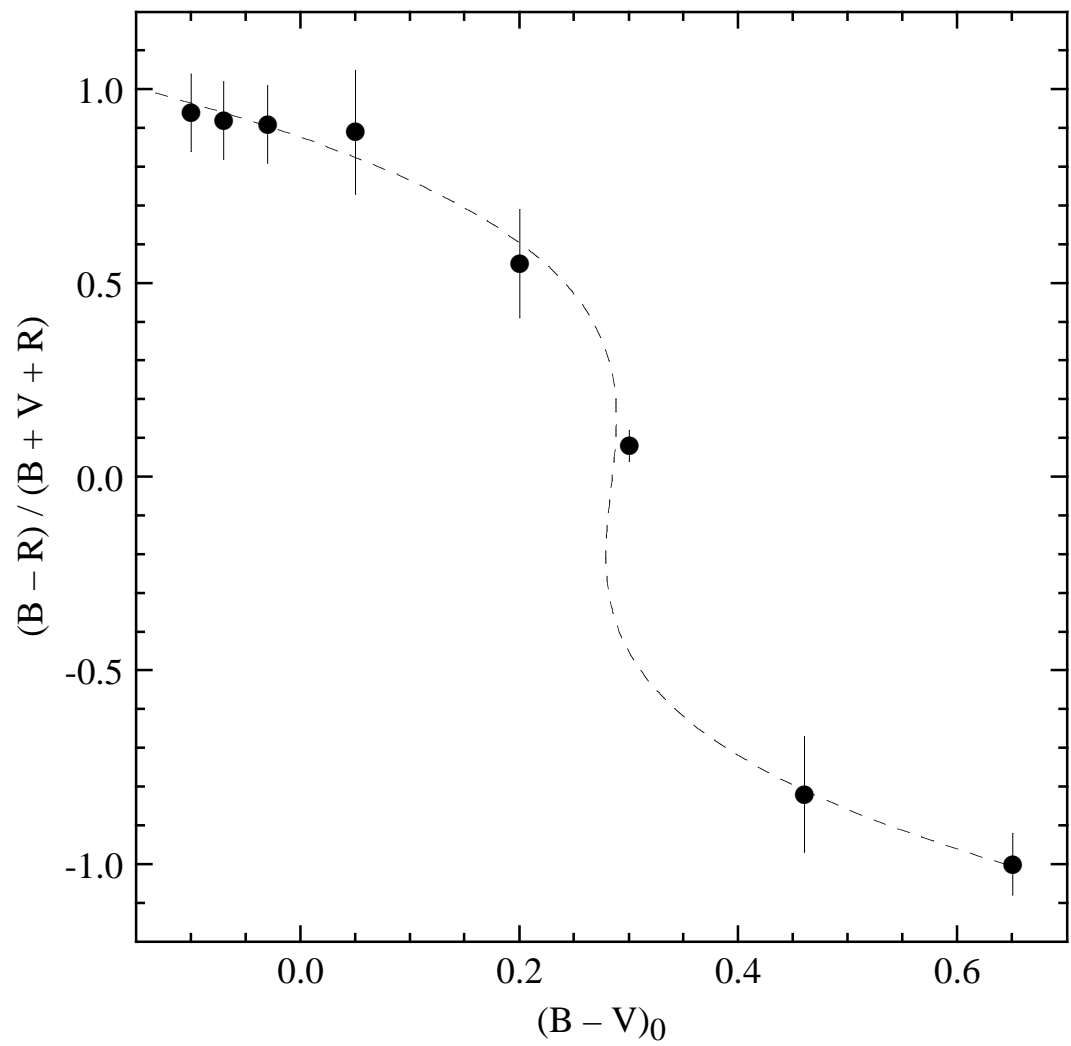
\end{document}